\documentstyle[preprint,prb,aps]{revtex}
\begin{document}
\draft
\title{ {\it Phys. Rev. B, \, \,  Rap. Comm. (to appear) } \\
Decoupling in the 1D frustrated quantum XY model and Josephson
junction ladders:  Ising critical behavior }

\author{Enzo Granato \\
Laborat\'orio Associado de Sensores e Materiais, \\
Instituto Nacional de Pesquisas Espaciais, \\
12225 S\~ao Jos\'e dos Campos, S.P. Brazil}

\maketitle

\begin{abstract}
A generalization of the one-dimensional frustrated quantum XY model is
considered  in which the inter and intra-chain  coupling  constants of
the two  infinite  XY (planar rotor) chains  have different strengths.
The model can describe the superconductor to insulator transition due to
charging effects in a ladder of Josephson junctions in a magnetic field
with half a flux quantum per plaquette. From a fluctuation-effective
action, this transition is expected to be in the universality class of
the two-dimensional classical XY-Ising model. The critical behavior  is
studied using a Monte Carlo transfer matrix applied to the
path-integral representation of the model and a finite-size-scaling
analysis of data on small system sizes.  It is found that, unlike the
previous studied case of equal inter and intra-chain coupling
constants, the XY and Ising-like excitations  of the quantum model
decouple for large interchain coupling, giving rise to pure Ising model
critical behavior for the chirality  order parameter and a
superconductor-insulator transition in the universality class of the 2D
classical XY model.
\end{abstract}
\newpage

The one-dimensional frustrated quantum XY model  (1D FQXY) has been
introduced as a model for studying charging effects in a ladder of
Josephson junctions in a magnetic field corresponding to  half  a
flux quantum per unit cell \cite{Granato}. These charging effects arise from
the
small capacitance of the grains making up the ladder and leads to strong
quantum fluctuations of the phase of the superconducting order
parameter. As a result of the competition between  the charging energy
and the Josephson coupling between the superconducting grains, the
one-dimensional array undergoes a superconductor to insulator
transition at zero temperature for decreasing capacitance
\cite{Geerligs,vdZant,Physica}.
The universality class of this transition is currently a problem of great
interest specially in relation to experiments on two-dimensional
superconducting films and Josephson-junction arrays
\cite{Geerligs,vdZant,Haviland,Fisher,GranatoK90}.  For a chain of
Josephson junctions this critical behavior has been identified with
that of the well known classical two-dimensional XY model \cite{Doniach}.
However, while for the case of two coupled chains forming a ladder, in the
absence of a magnetic field, this critical behavior  can be shown to
remain unchanged, in the presence of a magnetic field the
behavior is
more complicated\cite{Granato,Kardar}. In particular, at half flux quantum per
plaquette,
corresponding to the 1D FQXY model \cite{foot1} we study here, the critical
beha
   vior
is expected to be described by the 2D classical XY-Ising model
\cite{Granato,GranatoK91}.
The existence of both XY and Ising-like excitations in this case result
from the frustration induced by the magnetic field and are associated
with the continuous $U(1)$ symmetry of the phases of the
superconducting order parameter and the plaquette chiralities which
measures the direction of circulating currents in the
Josephson-junction ladder.

The 1D FQXY model is defined by the
Hamiltonian \cite{Granato}
\begin{equation}
H = -{{E_c}\over 2} \sum_r \left( {d \over { d \theta_r }} \right) ^2
- \sum_{<rr^\prime>} E_{rr^\prime} \cos ( \theta_r - \theta_{r^\prime})
\end{equation}
and consists of a one-dimensional chain of frustrated plaquettes as
indicated in Fig. 1.  The first term in Eq. (1) describes quantum
fluctuations induced  by the charging energy  $E_c = 4 e^2/C$ of a
non-neutral superconducting grain located at site $r$, where $e$ is the
electronic charge and $C$ is the effective capacitance  of the grain.
The second term is the usual Josephson-junction coupling between
nearest-neighbor grains. $\theta_r$ represents the phase of the
superconducting order parameter  and the couplings $E_{rr^\prime}$
satisfy the Villain's 'odd rule' in which the number of negative bonds
in an elementary cell is odd \cite{Villain}. This rule is a direct
consequence of
the constraint that, for the half flux case, the line integral of the
vector potential due to the applied magnetic field should be equal to $
\pi$ in units of the flux quantum \cite{foot2}. In the classical limit ($E_c =
0$), the ground state of Eq. (1) has a discrete $Z_2$ symmetry associated
with an antiferromagnetic pattern of plaquette chiralities $\chi_p =
\pm 1$ measuring the two opposite directions of the super-current
circulating in each plaquette. In a previous  work \cite{Granato}, a
particular
case of the 1D FQXY model, i.e., $ E_y  = E_x $,  has been studied in some
detail and it has been found  that in fact its critical behavior is
consistent with the results for the 2D classical XY-Ising model
\cite{GranatoK91}. From
the critical exponents associated to the chirality order parameter the
critical behavior has been identified as the one along the line of
single transitions where both phase coherence and chiral order are lost
simultaneously.  However, the XY-Ising model has in addition to this
transition line, two other branches corresponding to separate XY and
Ising critical behavior which join the line of single transition at a
bifurcation point located at some place in  the phase diagram. The 1D
FQXY model studied previously corresponds to a particular path through
this phase diagram; the one located in the region of single
transitions.

In this work we consider a generalized  version of the 1D FQXY model in
which the inter ($E_x$) and intra-chain  ($E_y$) couplings constants
have different strengths. In terms of a fluctuation-effective action
which can obtained from an imaginary-time path integral representation
of  Eq. (1), the ratio between the couplings  constants $ E_x/E_y$ can
be used to tune the system through the bifurcation point in the
XY-Ising model \cite{Granato}.  In particular, for $E_x >> E_y$ the 1D FQXY
model is
expected to  have two separate transitions. In this work we find that
in fact for $E_x/E_y \sim 2$  the single transition found previously
does decouple into two separated transitions. Using a Monte Carlo
transfer matrix technique \cite{Nightingale} applied to the path-integral
representation
of the model we study the critical behavior of the chirality order
parameter at a particular value of this ratio, $E_x/E_y=3$. We find,
from   a finite-size scaling analysis of extensive calculations on
small system sizes, that the  critical exponents are  consistent with
pure Ising model critical behavior as expected from the results for the
XY-Ising model. Thus the superconductor to insulator transition in the
related  Josephson junction ladder is in the universality class of
the 2D classical XY model.

To study the critical behavior of the 1D FQXY model, we find it
convenient to use an imaginary-time path-integral formulation of the
model \cite{ZinnJustin}.  In this formulation, the one-dimensional quantum
problem maps into a 2D classical statistical mechanics problem where the
ground state energy of the quantum model of finite size $L$ corresponds
to the reduced free energy per unit length of the classical model
defined on an infinite strip of width $L$ along the imaginary time
direction, where the time axis $\tau$ is discretized in slices $\Delta
\tau$. After scaling the time slices appropriately in order to get a
space-time isotropic model, the resulting classical partition function
is given by $Z = tr e^{-H}$ where the reduced classical Hamiltonian is
defined as
\begin{eqnarray}
H = -\alpha \sum_{\tau,j}\, \, [&
\cos(\theta_{\tau,j}-\theta_{\tau,j+1}) +
\cos(\theta_{\tau,j} - \theta_{\tau+1,j}) \cr
& \, - \cos(\phi_{\tau,j}-\phi_{\tau,j+1})+
\cos(\phi_{\tau,j} - \phi_{\tau+1,j}) \cr
&  + {{E_x}\over{E_y}} \cos(\theta_{\tau,j} - \phi_{\tau,j} ) \, \,]
 \end{eqnarray}
In the above equation, $\theta$ and $\phi$ denote the phases on the
left and right columns in Fig. 1, and $\alpha = (E_y/E_c)^{1/2}$ plays
the role of an inverse temperature in the 2D classical model.

One can now carry out a detailed study of the scaling behavior of the
energy gap for kink excitations (chiral domain walls) of the 1D FQXY
model by noting that this corresponds to the interface free energy of
an infinite strip in the model of Eq. (2).  For large $\alpha$ (small
charging energy $E_c$), there is a gap for creation of kinks in the
antiferromagnetic pattern of $\chi_p$ and the ground state has
long-range chiral order. At some critical value of $\alpha$, chiral
order is destroyed by kink excitations, with an energy gap vanishing as
$| \alpha -\alpha_c | ^\nu$, which defines the correlation length
exponent $\nu$. Right at this critical point, the correlation function
decays as a power law $ <\chi_p \chi_{p^\prime} > = |p-p^\prime|
^{-\eta}$ with a critical exponent $\eta$.  However to proceed further,
one has to be able to calculate the free energy  per unit length
$f(\alpha)$ of the Hamiltonian  on the infinite strip, which is
usually obtained from the largest eigenvalue $\lambda_o$ of the
transfer matrix between different time slices as $f=- \ln \lambda_o$.
Here, the major difficulty in performing this type of calculation comes
from the continuous degrees of freedom of the model which prevents an
exact diagonalization of the transfer matrix. This can be overcome by
using a Monte Carlo transfer-matrix method \cite{Nightingale} which has
been shown to
lead to accurate estimates of the largest eigenvalue even for this type of
problems.
Here we just summarize the main steps. The method is a stochastic
implementation of the well-known power method to obtain the dominant
eigenvalue of a matrix. First helical boundary conditions are
implemented, in order to get a sparse matrix. Then, a sequency of
random walkers $R_i$, $1 \le i \le r$, representing the configurations
of a column with $L$ spins in the infinite strip is introduced with
corresponding weights $w_i$. The number of walkers $r$ is maintained
within a few percent of a target value $r_o$ by adjusting the weights
properly. A matrix multiplication can be regarded as a transition
process with a probability density defined from the elements of the
transfer matrix. In the procedure, a Monte Carlo  (MC) step consists of
a complete sweep over all random walkers and after a large number of MC
steps an estimate of the largest eigenvalue can be  obtained from the
ratio between the total weights $\sum_i w_i$ of two successive MC
steps. The implementation and some of the difficulties of the method
are similar to the case of the two-dimensional frustrated classical XY
model \cite{GranatoN} and the  reader should refer to that work for further
details.
For the calculations discussed in this work, typically $r_o=20000$
random walkers and $80000$ MC steps were used which correspond to
$1.5\times 10^8$ attempts per $(\theta,\phi)$ pair.

The interfacial energy for domain walls in the model of Eq. (2) can be
obtained from the differences between the free energies for the
infinite strip with and without a wall. However, because of the
antiferromagnetic pattern of the chiralities $\chi_p = \pm 1$, only
strips with an odd number of sites $L$ will have a domain wall. Since
one is required to obtain the free energy differences at the same value
of $L$, we need to resort to an interpolation scheme for successive
odd  or  even $L$ to determine the interfacial free energy. Results for
the interfacial free energy, defined as $\Delta F(\alpha, L) = L^2
\Delta f(\alpha, L)$, near the transition point $\alpha_c$, for $6 < L
< 14$, are indicated in Fig. 2 for a particular value of the ratio
$E_x/E_y= 3$. As in the previous work \cite{Granato}, to obtain the critical
exponents and critical temperature we now employ the finite-size scaling

\begin{equation}
\Delta F(\alpha, L) = A(L^{1/\nu} \delta \alpha)
\end{equation}
where $A$ is a scaling function and $\delta = \alpha - \alpha_c$. In a linear
ap
   proximation for the argument of $A$, we have
\begin{equation}
\Delta F(\alpha, L) = a+ b  L^{1/\nu} \delta \alpha
\end{equation}
which can be used to determine the critical coupling $\alpha_c$ and the
exponent $\nu$ independently. The change from  an increasing trend with
$L$ to a decreasing trend provides and estimate of $\alpha_c$, which
from Fig. 2 gives $\alpha_c = 1.16 (2)$.  Once  the critical coupling
is known, the correlation function exponent $\eta$ can obtained from
the universal amplitude  $a$  in Eq. (4) through a result from
conformal invariance \cite{Cardy}, $a=\pi \eta$, from which we estimate
$\eta = 0.27(3)$. To estimate the correlation length exponent $\nu$ we
first obtain
the derivative $S= \partial \Delta F / \partial \alpha$  near
$\alpha_c$, then it can easily be seeing that a log-log plot of $S $ vs
$ L$ gives an estimate of $1/\nu$ without requiring a precise
determination of $\alpha_c$. Of course, this is only valid for the
linear approximation of Eq. (4).  Assuming the data is in fact in this
regime we obtain the result for $S$ as indicated in Fig. 3 and  get the
estimate $\nu= 1.05 (6)$. The results for the critical exponents $\nu$
and $\eta$ are good agreement with pure 2D Ising values $\nu =1$ and
$\eta=0.25$ indicating pure Ising behavior. Moreover, this implies from
the relation between the 1D FQXY model and the 2D classical XY-Ising
\cite{GranatoK91} that the XY and Ising-like excitations have decoupled in this
region.

To show that in fact one has two decoupled and at the time separated
transitions we now consider  the results for the helicity modulus which
measures the response of the system to an imposed phase twist. In the
incoherent phase this quantity should vanishes while it should be
finite in the coherent phase. The helicity modulus is  related to the
free-energy differences $\Delta F$ between strips with and without and
additional phase mismatch of $\pi$ along the strip and is given by
$\gamma = 2 \Delta F/\pi^2$ for large system sizes. If the model is
decoupled then  the coherent to incoherent (or
superconductor-insulator) transition should be in the universality class
of the 2D XY model, where one knows that the transition is associated
with a universal jump of $2 \pi$ in the helicity modulus \cite{Nelson}.
Finite-size
effects smooth out this behavior  as indicated in Fig 4 but the
critical coupling can be estimated as the value of $\alpha$ at which
$\Delta F = \pi$. This criteria leads to the estimate $\alpha_c=1.29$
which is to be compared with the critical coupling for the destruction
of chiral order found above, $\alpha_c = 1.16$. This clearly indicates
the transitions are well separated and thus one expects they are
decoupled.  In contrast, for the case of equal inter and intra-chain
coupling constants studied previously these estimates were found to be
in fact consistent with each other \cite{Granato}. We have also performed
less detailed  calculations at other values of the ratio $E_x/E_y$
from which we can estimate that the Ising and XY transition merge into
a single transition roughly at $E_x/E_y \sim 2$. Since, the superconductor
to insulator transition is to be identified with
the loss of phase coherence \cite {Doniach} we reach the interesting result
that in the 1D FQXY, or alternatively, a Josephson-junction ladder, the
universality class of the superconductor-insulator transition depends on the
ratio between inter and intra-chain couplings.

In conclusion, we have studied a generalized version of the
one-dimensional frustrated quantum XY model which consisted in allowing
for different strengths for the inter and intra-chain couplings
constants. The model can be physically realized as a  one-dimensional
array of Josephson junctions in the form of a ladder and in the presence
of an external magnetic field corresponding to a half flux quantum per
plaquette. It is found that, unlike the previous studied case of equal
inter and intra-chain couplings, the XY and Ising-like excitations decouple
giving rise to pure Ising behavior for chirality order parameter and a
superconductor-insulator transition in the universality class of the XY model.
Since these arrays can currently be fabricated in any
desired geometry and with well-controlled parameters it is hoped that
these results will serve to motivate experiments in these systems.

\section{ Acknowledgments}

The author would like to thank Prof. Abdus Salam, the International
Atomic Energy Agency and UNESCO for hospitality at the International
Centre for Theoretical Physics, Trieste, Italy, where this work was
completed. This work has been supported in part by Funda\c c\~ao
de Amparo \`a Pesquisa do Estado de S\~ao Paulo (FAPESP, Proc. no.
92/0963-5) and Conselho Nacional de Desenvolvimento Cient\'ifico e
Tecnol\'ogico (CNPq).

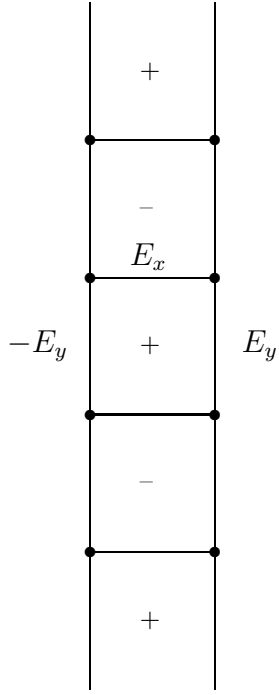
\begin{figure}
\caption{ Schematic representation of the one-dimensional frustrated
quantum XY
model with inter ($ E_x$) and intra-chain ($\pm E_y$) coupling constants.
The antiferromagnetic ordering of chiralities $\chi_p = \pm 1$ is also
indicated. }
\end{figure}
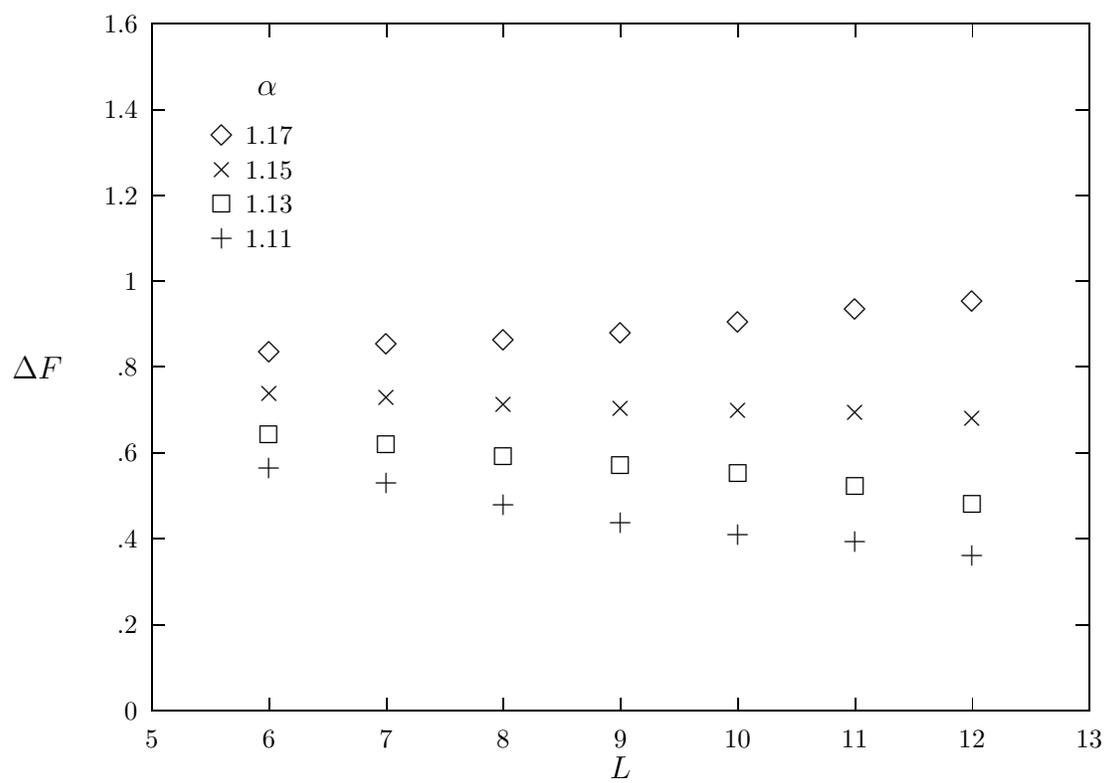
\begin{figure}
\caption{ Finite-size scaling of the interfacial free energy $\Delta F(\alpha,
L) = L^2 \Delta f(\alpha, L)$ for kink (chiral) excitations.}
\end{figure}
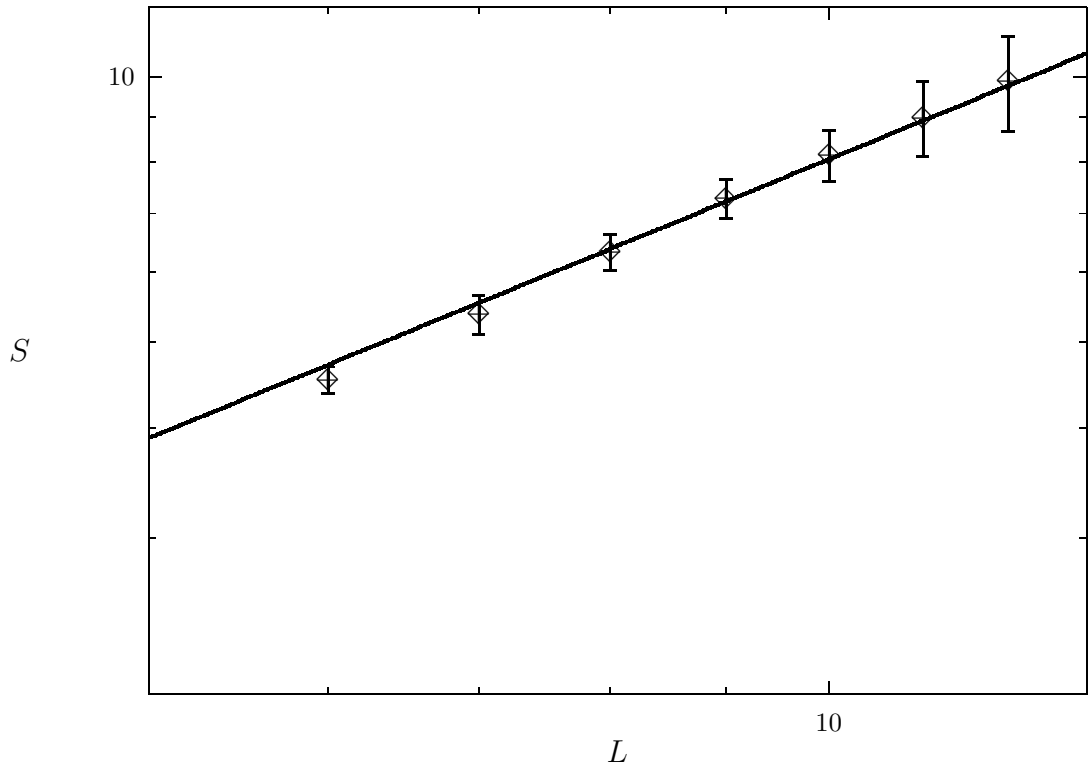
\begin{figure}
\caption{ $ S = \partial \Delta F(\alpha, L) / \partial \alpha$ evaluated near
the critical
coupling $\alpha_c$. The slope of the straight line gives an estimate of
$1/\nu$. }
\end{figure}
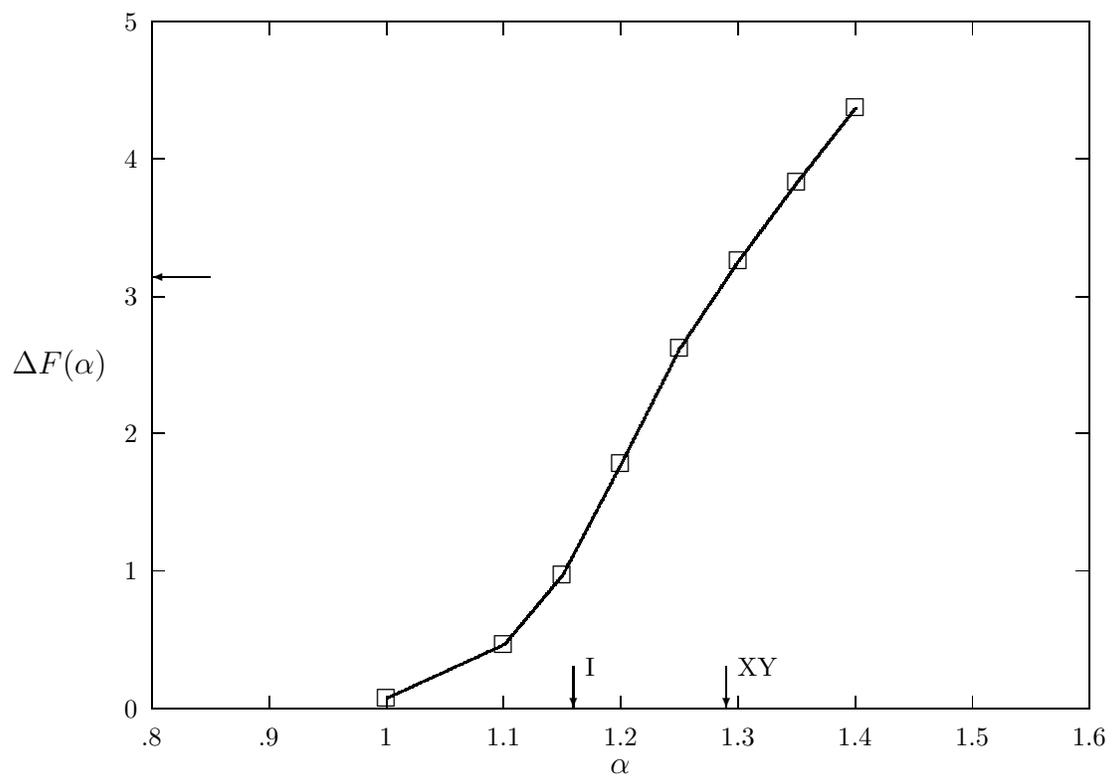
\begin{figure}
\caption{ Behavior of the interfacial free energy $\Delta F = L^2 \Delta f$
for a system of size $L=12$ resulting from an imposed  phase twist of $\pi$.
Vertical arrows indicate the locations of the Ising and XY transitions and
the horizontal arrow the value $\Delta F =\pi$
from where the XY transition is located.
The Ising transition is located from  the finite-size scaling of the
chiral order parameter (Fig. 2) as discussed in the text.}
\end{figure}

\newpage
\center
\setlength{\unitlength}{0.240900pt}
\ifx\plotpoint\undefined\newsavebox{\plotpoint}\fi
\sbox{\plotpoint}{\rule[-0.175pt]{0.350pt}{0.350pt}}%

\vfill Fig. 4
\end {center}


\begin{references}
\bibitem{foot1} In the case of a Josephson-junction chain \cite{Doniach}
the XY critical behavior follows from the equivalence between this model
Hamiltonian and the quantum Hamiltonian limit of the transfer matrix of
the classical 2D XY model. More generally, the  critical behavior of a
$d$ dimensional quantum model is in the same universality class of the
$d+1$ dimensional classical version \cite{ZinnJustin}.  The 1D FQXY
model, apparently, is not the Hamiltonian limit of  the 2D classical
model, i.e., the 2D frustrated XY model. Yet, they  critical behavior
appear to be in the  same universality class \cite{Granato,GranatoK91}.
\bibitem{foot2} The second term in Eq. 1 arises from the Josephson
coupling between
the grains which has the form
$\sum_{<ij>} J_{<rr^\prime>} \cos (\theta_r -
\theta_{r^\prime}-A_{rr^\prime}) $ where $J_{rr^\prime} > 0$ and
$A_{rr^\prime} = (2\pi/\Phi_o) \int_r ^{r^\prime} \vec A \cdot d l$ is
the line integral of the vector potential $\vec A$. Since $\nabla \vec
A = \vec B$, one has the constraint $\sum A_{rr^\prime} = 2\pi f$
around an elementary plaquette, where $f$ is the number of flux quanta
per plaquette. For the fully frustrated case $f=1/2$, using a   Landau
gauge where $\vec A = xB \hat y$ leads to antiferro and ferro-magnetic
bonds as indicated  in Fig. 1.
\end{references}
\end{document}